\documentclass[aps,prl,10pt,twocolumn,superscriptaddress]{revtex4}
\usepackage{graphicx}

\def\ET{\rm BEDT-TTF}
\def\ETI{({\ET})$_2$I$_3$}
\def\AETI{$\alpha$-{\ETI}}
\def\BETI{$\beta^*$-{\ETI}}
\def\KETI{$\kappa$-{\ETI}}
\def\k{{\bf k}}                 % Bold k
\def\q{{\bf q}}                 % Bold q

\begin{document}

\title{BEDT-TTF organic superconductors: the entangled role of phonons}

\author{Alberto Girlando}
\author{Matteo Masino}
\affiliation{Dip. Chimica Gen.Inorg. Chim.Anal. e Chim.Fis. and
INSTM-UdR Parma, Parma University, 43100 Parma, Italy}
\author{Aldo Brillante}
\author{Raffaele G. \surname{Della Valle}}
\author{Elisabetta Venuti}
\affiliation{Dip. Chimica Fisica ed Inorganica
and INSTM-UdR Bologna, Bologna University, Bologna, Italy}

\date{\today}

\begin{abstract}
We calculate the lattice phonons and the electron-phonon coupling of the
organic superconductor {\KETI}, reproducing all available experimental data
connected to phonon dynamics. Low frequency intra-molecular vibrations are
strongly mixed to lattice phonons. Both acoustic and optical phonons are
appreciably coupled to electrons through the modulation of the hopping
integrals ({\it e}-LP coupling). By comparing the results relevant to
superconducting $\kappa$- and {\BETI}, we show that electron-phonon coupling
is fundamental to the pairing mechanism. {\it Both} {\it e}-LP and
electron-molecular vibration ({\it e}-MV) couplings are essential to
reproduce the critical temperatures. The {\it e}-LP coupling is stronger,
but {\it e}-MV is instrumental to increase the average phonon frequency.
\end{abstract}

\maketitle

Organic superconductors (oSC) have been studied for more than twenty years,
but the nature of the pairing mechanism is still unknown, and remains one of
the most intriguing problems in this class of materials \cite{review}. Like
high $T_c$ superconductors, oSC are low-dimensional systems characterized by
strong electronic correlations. This fact and the proximity of the
superconducting phase to magnetically or charge ordered states has brought
about the suggestion of exotic pairing, mediated for instance by spin or
charge fluctuations \cite{exotic}. However, direct experimental evidence for
such kind of purely electronic mechanisms has not been provided. On the
other hand, the oSC phonon structure is very complex, preventing a clear
assessment of the role of phonons in the pairing. Indeed, scattered
experimental evidences of the involvement of intra-molecular phonons
\cite{merzhanov}, of optical inter-molecular phonons \cite{rama-eph}, and of
acoustic phonons \cite{pinto} have been presented. However, all these
studies, focusing on a particular type of phonon, failed to offer a
comprehensive and convincing picture of the phonon mediated mechanism in
oSC. Aim of this letter is to provide a unified framework for the
description of the phonon dynamics and of the electron-phonon coupling in
oSC based on bis-ethylendithio-tetrathiafulvalene ({\ET}). We shall show
that all the above types of phonons are coupled to electrons, and cooperate
in establishing electron pairing.

In molecular crystals it is convenient to start by distinguishing between
intra- and inter-molecular vibrational degrees of freedom, and to consider
their interactions at a later stage. Thus, in the framework of a
tight-binding description of the electrons, intra-molecular vibrations
modulate the on-site energy and yield the electron-molecular vibration ({\it
e}-MV) coupling, while lattice phonons, mainly inter-molecular in character,
modulate the hopping integrals ({\it e}-LP coupling).

Molecular vibrations and {\it e}-MV coupling constants are transferable
among different crystalline phases, and either computational or experimental
methods can be used for their determination. Indeed, {\ET} molecular normal
modes and {\it e}-MV coupling constants are fairly well known \cite{emv}.
The problem of lattice phonons is much harder to deal with, because the
experimental characterization of lattice phonons and of {\it e}-LP coupling
constants is difficult and subject to considerable experimental uncertainty
\cite{rama-eph}. Moreover, the phonon properties are not transferable among
different crystals. To overcome these difficulties, we have adopted the
``Quasi Harmonic Lattice Dynamics'' (QHLD) computational method
\cite{beta,QHLD}, which is known to accurately predict the crystal structure
and the phonon dynamics of complex molecular crystals as a function of
pressure $p$, and temperature $T$.

As extensively discussed elsewhere \cite{beta,QHLD}, in QHLD the vibrational
contribution to the crystal Gibbs energy $G(p,T)$ is approximated by the
free energy of the phonons calculated in the harmonic approximation. The
structure at a given $p$,~$T$ is determined self-consistently by minimizing
$G(p,T)$ with respect to lattice parameters, molecular positions and
orientations. The method relies on an accurate inter-molecular potential
model, which we have described as the sum of empirical atom-atom potentials,
known to be reasonably transferable among crystals containing similar
molecules. We have restricted our attention to the {\ETI} salts, which
constitute an ideal benchmark for our approach. Our strategy has been to
determine and to fix once and for all the needed C, S, H and I atom-atom
parameters on the basis of the crystal structures of neutral {\ET} and
non-superconducting {\AETI} salt \cite{QHLD}. We have then applied the
method to the phonon structure and {\it e}-LP coupling strength of
superconducting {\BETI} salt \cite{beta}. Here we illustrate the
corresponding results for {\KETI}. The comparison with the experimental data
for the two superconducting $\beta^*$- and $\kappa$-phases allows us to
validate our approach and to assess the role of phonons in the
superconductivity mechanism.

Since both I$_3^-$ and {\ET} present low frequency intra-molecular modes
falling in the region of the lattice phonons, the coupling between inter-
and intra-molecular phonons may be important for the {\ETI} salts. To
account for this effect, we use the exciton-like model described elsewhere
\cite{beta}. We stress that lattice phonons are actually mixed only with
low-frequency vibrations (below 300 cm$^{-1}$), and these do not modulate
on-site energies \cite{emv}. Therefore {\it e}-LP and {\it e}-MV couplings
can be dealt with separately.
\begin{figure}
\includegraphics*[scale=0.4]{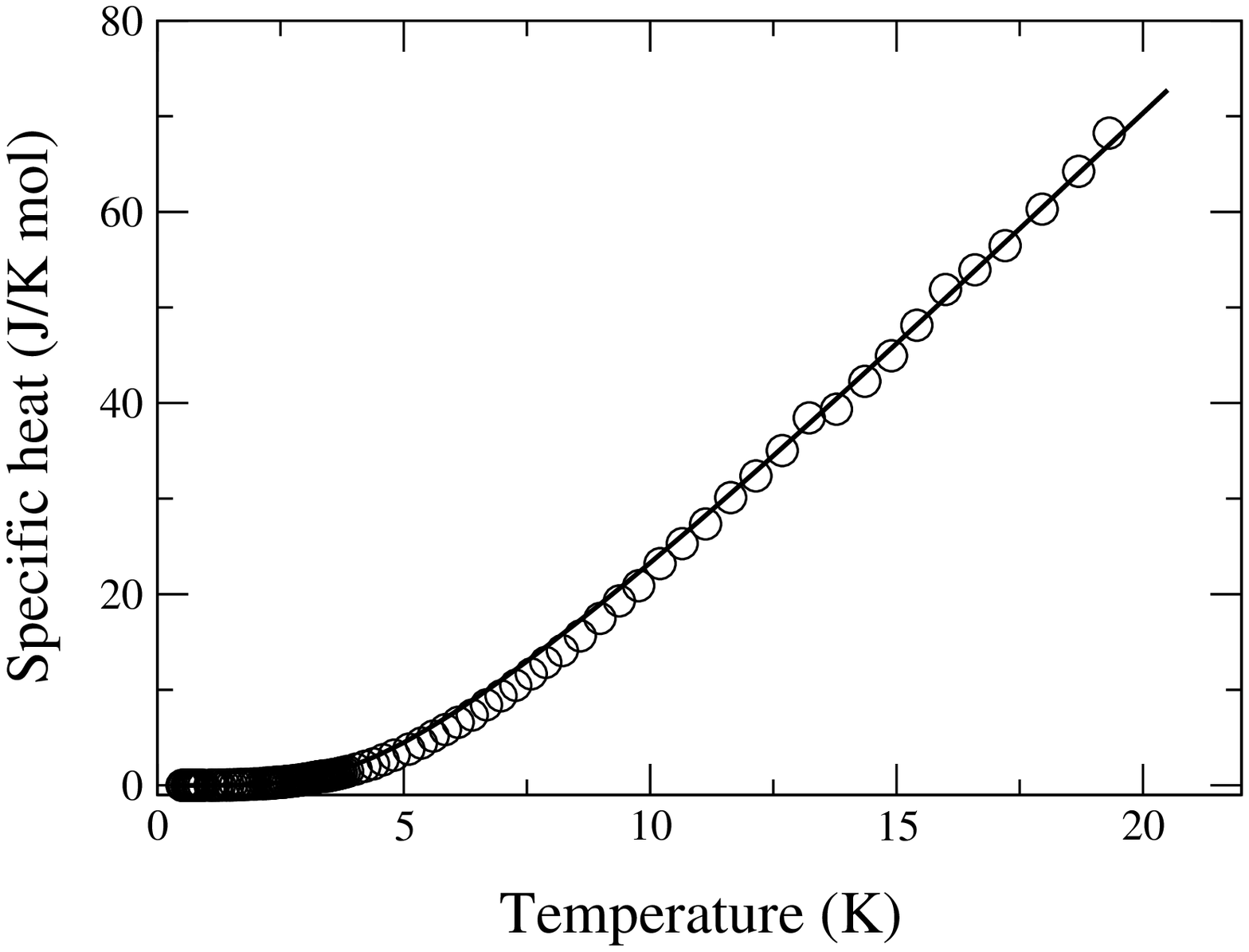}
\caption{\label{s_heat} Calculated ($c_V$, full line) and experimental
($c_p$, circles, from Ref.~\protect\onlinecite{wosnitza}) {\KETI} specific
heat.}
\end{figure}

At room temperature, {\KETI} crystallizes in the monoclinic system
$P2_{1}/c$, with four molecules per unit cell. Upon cooling below 150 K the
crystal structure changes to the $P2_{1}$ system \cite{japastruct}. The
{\ET} layer is in the $bc$ plane. The QHLD calculations reproduce both
crystal structures, which correspond to two distinct $G(p,T)$ minima. The
calculated lattice parameters are within 3$\%$ of the experimental ones. The
results presented below all refer to the low temperature $\kappa$-phase.

A comparison with {\KETI} scanty infrared and Raman data does not offer a
stringent test for the calculated phonon frequencies, whose number largely
exceeds that of the currently observed bands. We then compare the
experimental \cite{wosnitza} specific heat $c_p$ to the calculated $c_V$,
which depends on the overall distribution of phonon frequencies $\omega_{\q
j}$: $c_V(T) = k_{\rm B} \sum_{\q j} F(\hbar\omega_{\q j}/k_{\rm B} T)$,
where $F(x) = x^2 {\rm e}^x ({\rm e}^x\!-\!1)^{-2}$ and the sum is extended
to all phonon branches $j$ and wavevectors ${\q}$. As shown in
Fig.~\ref{s_heat}, the agreement is outstanding. We stress that there are
{\it no adjustable parameters} in our $c_V(T)$ calculation.

\begin{figure}
\includegraphics*[scale=0.45]{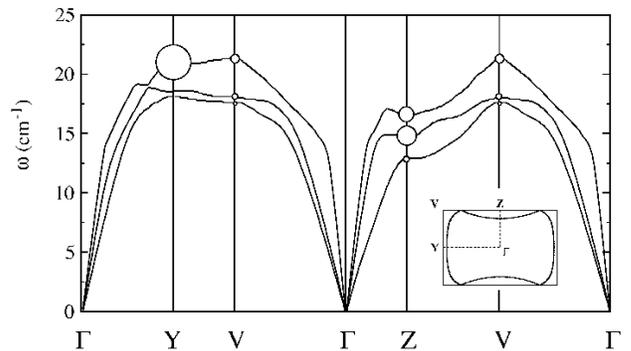}
\caption{\label{dispersion} Dispersion curves of {\KETI} acoustic lattice
phonons along the principal directions shown in the inset. Inset: Fermi
surface calculated with the dimer model.}
\end{figure}

To calculate the {\it e}-LP coupling strength we assume that the low
frequency phonons, mainly intermolecular in character, couple to electrons
only through the modulation of hopping integrals:
\begin{equation}
t_{KL} = t^0_{KL} + \sum_j g(KL;{\q} j)~ Q_{{\q} j}
\end{equation}
where $t_{KL}$ is the hopping integral between neighboring pairs $KL$ of
{\ET} molecules, and $Q_{{\q} j}$ are the dimensionless normal
coordinates. Finally, $g(KL;{\q} j) = \sqrt{\hbar/2\omega_{\q j}}(\partial
t_{KL}/\partial Q_{{\q} j})$ defines the linear {\it e}-LP coupling
constants. The $t^0_{KL}$ and their dependence from the QHLD eigenvectors
$Q_{{\q} j}$ have been calculated by the extended H\"uckel (EH) method
\cite{beta}. The obtained $t^0$ values are in good agreement with those
calculated by extended basis set density functional methods \cite{tdensfun}.
The $t^0$ are used to describe the band structure in a tight binding
approach, adopting the dimer model \cite{dimer}. The resulting conduction
bands and Fermi surface (inset of Fig.~\ref{dispersion}) match closely those
obtained considering all the four site orbitals within the unit cell
\cite{japastruct}.

To express the {\it e}-LP coupling constants in reciprocal space, we have to
introduce the dependence on both electronic ({\k}) and phononic ({\q})
wavevectors. We have followed the approximation scheme adopted for {\BETI}
\cite{beta}. The {\k} dependence has been accounted for in terms of the
dimer model: the coupling constants associated with the modulation of the
intra-dimer hopping integral are assumed independent of {\bf k}, whereas the
inter-dimer {\rm g}'s are assumed to have the following dependence: ${\rm
g}({\k,\k}';j) = \sum_{\bf R}g_{KL}({\q} j)~ ({\rm e}^{i{\k}'{\bf R}}-{\rm
e}^{i{\k}{\bf R}})$, where ${\q} = {\k}'- {\k}$, and {\bf R} represents the
nearest neighbor lattice vectors. For what concerns the {\q} dependence, the
optical modes are taken independent of ${\q}$, whereas for acoustic phonons
we have simply assumed $g_{KL}({\q} j) \propto \sqrt{\mid {\q}\mid}$ up to
the average value calculated at the zone edges.

Fig.~\ref{dispersion} shows the frequency dispersion of the acoustic modes
along representative Brillouin zone directions, together with a pictorial
representation of the coupling strength at the zone edges. The dots
diameters are proportional to the squared coupling constant, which are much
larger along the Y, Z directions ({\it i.e.} the $b^*$, $c^*$ axes) than
along the diagonal. Our results indicate that for {\KETI} the coupling of
acoustic phonons is very anisotropic.
\begin{figure}
\includegraphics*[scale=0.4]{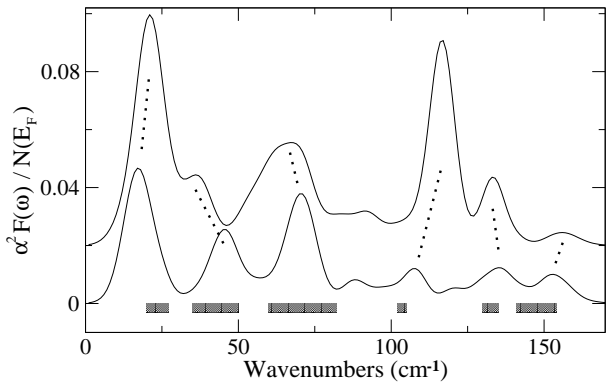}
\caption{\label{elias} Calculated Eliashberg function of {\KETI} (lower
curve) and {\BETI} (upper curve, offset for clarity). The shaded areas at
the bottom indicate the position of the most strongly coupled phonons, as
detected from Raman on others $\kappa$-phase {\ET} salts (from
Ref.~\protect\onlinecite{rama-eph}).}
\end{figure}

From the coupling constants in reciprocal space and the QHLD phonon density
of states we calculate the Eliashberg coupling function \cite{allen}:
$\alpha^2(\omega)F(\omega) = N(E_F)\sum_{j}\langle|{\rm g}(\k,\k';j)|^2
\delta(\omega-\omega_{\q j})\rangle_{FS} $, where $N(E_F)$ is the unit cell
density of states/spin at the Fermi level, and $\langle\,\, \rangle _{FS}$
indicate the average over the Fermi surface. Fig.~\ref{elias} compares
$\kappa$- and {\BETI} Eliashberg functions. Both curves have been smoothed
by a Gaussian of constant width to ease the comparison. Although significant
differences exist, also due to the higher number of phonons for the
$\kappa$-phase unit cell, the most significant peaks are correlated, as
indicated by the dotted lines. In both salts the lowest frequency peak
results from the overlap of acoustic and lowest frequency optical modes.

Comparison of the computed Eliashberg function with available experimental
data provides a test for the {\it e}-LP constants and QHLD eigenvectors. For
the $\beta$ phase, the calculation compares quite favorably \cite{beta} with
the Eliashberg function directly derived from point contact measurements.
This kind of measurement is not available for {\KETI}, but we can make
reference to the phonon self-energy effects measured by Raman on
$\kappa$-({\ET})$_2$Cu(NCS)$_2$ and $\kappa$-({\ET})$_2$Cu[N(CN)$_2$]Br
\cite{rama-eph}. A quantitative comparison is not possible, due to the
different counter-ions. Nevertheless, Fig.~\ref {elias} shows that the
frequency regions (shaded areas at the bottom) of the few strongly coupled
optical phonons \cite{rama-eph}, indeed correspond to the peaks of {\KETI}
Eliahsberg function. We finally mention that inelastic neutron scattering
on $\kappa$-({\ET})$_2$Cu(NCS)$_2$ \cite{pinto} evidences also the
involvement of acoustic phonons, in agreement with our findings.

\begin{figure}
\includegraphics*[scale=0.4]{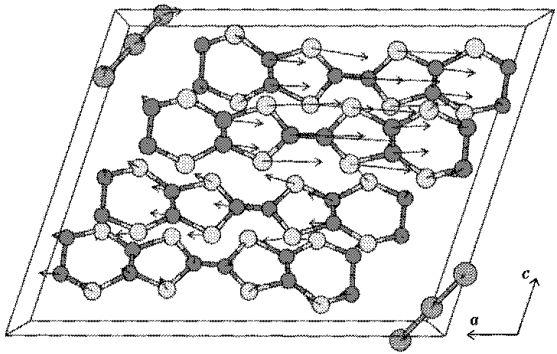}
\caption{\label{21eigen} Eigenvector of {\KETI} zone edge (Y) acoustic
phonon at 21 cm$^{-1}$.}
\end{figure}
Although further measurements, like Raman spectra with excitation in the
infrared, are highly desirable to provide further tests, we believe that the
above comparison already gives confidence to the QHLD eigenvectors, and as
an example we show in Fig.~\ref{21eigen} the eigenvector corresponding to
{\KETI} zone edge acoustic phonon at 21 cm$^{-1}$. As it occurs in several
other cases, the coupling is induced by the relative displacement of {\ET}
along the long molecular axis, {\it i.e.} roughly normal to the {\ET}
layers. Moreover, a normal mode coupled to electron is in general a mixture
of both {\ET} and I$_3$ vibrations. This mode mixing indicates that also
the counter-ion vibrations may play an indirect role in the coupling.

We now proceed to estimate the critical temperatures. From Fig.~\ref{elias}
Eliashberg function we derive the dimensionless coupling constant
$\lambda^{\rm LP}$, and the logarithmic average frequency $\omega_{\rm
ln}^{\rm LP}$ \cite{allen}:
\begin{eqnarray}
&\lambda^{\rm LP} = 2\int_0^{\omega_{\max}}
{\frac{\alpha^2(\omega)F(\omega)}{\omega}~ d\omega}
\\
&\omega_{\rm ln}^{\rm LP} =
\exp{\left(\frac{2}{\lambda_{\rm LP}}\int_0^{\omega_{\max}}
{\frac{\alpha^2(\omega)F(\omega)}{\omega}\ln\omega~ d\omega}\right)}
\end{eqnarray}
The $\lambda$ value depends on $N(E_F)$. With our tight binding EH
calculation we estimate $N(E_F)$ = 5.4 and 3.8 spin states/(eV unit cell),
so that $\lambda^{\rm LP}$ = 0.59 and 0.65 for $\kappa$- and {\BETI},
respectively. The corresponding $\omega_{\rm ln}^{\rm LP}$ values are 27
and 40 cm$^{-1}$. When these values are inserted into the Allen-McMillan
equation \cite{allen}:
\begin{equation}
T_c = \frac{\omega_{\rm ln}}{1.2}\exp\left [-\frac{1.04(1+\lambda)}
{\lambda - \mu^*(1+0.62\lambda)}\right ],
\end{equation}
assuming the standard 0.1 value for the Coulomb pseudopotential $\mu^*$, one
obtains $T_c$ = 0.8 and 1.7 K for $\kappa$- and {\BETI}, respectively, to be
compared with the experimental values of 3.4 and 8.1 K \cite{wosnitza,beta}.
Of course, the $T_c$ values depends critically on $N(E_F)$, a parameter
difficult to evaluate \cite{merino}. However, even by taking the maximum
current estimates, derived from experiment and therefore enhanced by many
body electron-electron and electron-phonon effects \cite{merino}, the $T_c$
values (2.7 and 3.6 K) remain significantly below the experiment. Therefore
{\it e}-LP coupling {\it alone} cannot reasonably account for the $T_c$ of
the two salts.

However, one must not forget high frequency {\it e}-MV coupled phonons
\cite{emv,pedron}. By including these phonons into the calculation of
$\lambda$ and $\omega_{\rm ln}$, the above scenario changes
substantially. We underline that {\it e}-MV coupling gives an additive
contribution to the total $\lambda$, $\lambda$ = $\lambda^{\rm
LP}+\lambda^{\rm MV}$, whereas the contribution to $\omega_{\rm ln}$ is {\it
multiplicative}: $\omega_{\rm ln} = \omega_{\rm ln}^{\rm LP}\omega_{\rm
ln}^{\rm MV}$, where $\omega_{\rm ln}^{\rm MV} = \exp\left [ \frac{2N(E_F)}
{N\lambda}\sum_l\frac{g_l^2}{\omega_l}\ln{\omega_l}\right ]$, with $g_l$ the
{\it e}-MV coupling constants and $N$ the number of molecules per unit cell
\cite{pedron}. It turns out that by including the {\it e}-MV coupling,
$\lambda$ increases by only $\sim$ 25 $\%$, but $\omega_{\rm ln}$ reaches 60
and 94 cm$^{-1}$ for $\kappa$- and {\BETI}, respectively. By using these
new values into Eq. (3), keeping $\mu^* =$ 0.1, we find that our tight
binding EH $N(E_F)$ estimates, which are within the range of the currently
accepted values \cite{merino}, account for the observed critical
temperatures. In particular, precise $T_c$ matching requires $N(E_F)$ = 5.2
and 3.9 spin states/(eV unit cell), corresponding to a total $\lambda$ of
0.74 and 0.91 for $\kappa$- and {\BETI}, respectively. We finally remark
that only by using the new $\omega_{\rm ln}$, which does not depend on
$N(E_F)$, the empirical relation \cite{marsiglio} connecting this parameter
to the specific heat jump $\Delta C_p/\gamma T_c$ yields for {\KETI} a jump
in nice agreement with experiment (1.7 {\it vs} 1.6) \cite{wosnitza}.

In our approach electron-phonon coupling provides a consistent and plausible
explanation of many experimental findings related to the superconducting
properties of $\kappa$- and $\beta^*$-phase {\ET} salts. Besides the just
mentioned $T_c$ values and specific heat jump, the scattered experimental
evidences of the involvement of acoustic and optical lattice phonons have
been rationalized. The coupling of acoustic phonons in the $\kappa$-phase
turns out to be strongly anisotropic, likely yielding an anisotropic gap
that might confuse the node search \cite{gap}. In addition, we have found
that the most strongly {\it e}-LP coupled {\ET} modes imply motions
approximately normal to the conducting planes, with counter-ions vibrations
mixed to {\ET} ones (cf. Fig.~\ref{21eigen}). This finding might well
account for the ``interlayer effects'' detected in thermal expansion
measurements \cite{muller}. Experimental evidences in favor of the
involvement of {\it e}-MV coupled phonons are mainly related to isotopic
effects \cite{merzhanov,schlueter}. In particular, the universal inverse
deuterium isotope effects (0.25 K) is likely due to a small increase in
$\omega_{\rm ln}$. The {\it e}-MV coupling constants of deuterated {\ET}
are not known, but we have verified that very small and plausible changes of
the coupling constants with respect to those of pristine {\ET} \cite{emv}
may indeed lead to the required increase in $T_c$, despite the large {\it
downwards} shift of several frequencies upon deuteration.

The principal conclusion of our paper is that phonons are the main
responsible for the coupling mechanism in {\ET} based oSC. The role of
phonons is complex, or ``entangled'', in the sense that different kinds of
phonons are all important, but contribute in different ways. The acoustic
and optical lattice phonons, modulating $t$, give the main weight to
$\lambda$, but intra-molecular phonons modulating on-site energies are
essential in increasing the average $\omega_{\rm ln}$ value. Although here
we vindicate the role of phonons, we also stress that traditional phonon
mediated pairing {\it alone} cannot give reason of all the experimental
findings related to superconducting properties of {\ET} salts. The proximity
of antiferromagnetic or charge ordered states in many oSC clearly indicate
that electron-electron interactions are important \cite{review}, and these
cannot be properly accounted for in our effective single particle
approach. In particular, just consider the above mentioned
$\kappa$-({\ET})$_2$Cu(NCS)$_2$ and $\kappa$-({\ET})$_2$Cu[N(CN)$_2$]Br
salts. Their $\omega_{\rm ln}$ is similar to {\KETI}, yet their $T_c$'s are
9.5 and 11.5 K, respectively \cite{elsinger}. We think it is unlikely that
the corresponding increase in $\lambda$ with respect to {\KETI} can be
attributed to a significant change in the {\it e}-LP coupling strength due
to the change in counter-ions. Rather, we believe that electron-electron
interactions and/or the proximity of a phase boundary may lead to an
enhancement of the effective electron-phonon coupling. In other words, we
need to analyze the role of electron-phonon coupling in low-dimensional
strongly correlated materials, still a poorly understood topic
\cite{lanzara}. Our work represents the necessary prerequisite for such an
analysis in {\ET}-based oSC.

\begin{acknowledgments}
Work supported by the Italian Ministero Istruzione, Universit\`a e Ricerca
(M.I.U.R.). We thank L. Frediani for the DFT calculations, and A. Painelli
for discussions.
\end{acknowledgments}
\end{document}